\documentclass[]{emulateapj}

\begin{document}

\title{Feedback and its Feedback Effect on Feedback: \\
Photoionization Suppression and its Impact on Galactic Outflows}

\author{Matthew M. Pieri\altaffilmark{1} and Hugo Martel\altaffilmark{1}}

\altaffiltext{1}{D\'epartement de physique, g\'enie physique et optique,
Universit\'e Laval, Qu\'ebec, QC, G1K 7P4, Canada}

\begin{abstract}

We show that radiative feedback due to reionization has a pronounced effect on the extent of mechanical feedback due to galactic outflows. The photoionization of the Intergalactic Medium (IGM) suppresses low-mass galaxy formation by photoheating the gas and limiting atomic line cooling. The number of low-mass galaxies is central for the enrichment of the IGM as these objects have the capacity to enrich a significant fraction (by volume) of the Universe. We use a modified version of our galactic outflow model, combined with a simple criterion for suppression, to investigate the potential impact upon the IGM. We find that this suppression strongly reduces the enrichment of the IGM and is sensitive to the reionization history. We also investigate the contribution of halos of different masses with varying degrees of suppression.
\end{abstract}


\section{INTRODUCTION}

Feedback is an essential component of our understanding and modeling of the intergalactic medium and galaxy formation. This feedback falls broadly into two categories: radiative feedback and mechanical feedback. They both act to deposit energy into the intergalactic medium (IGM) which in turn suppresses star formation. 

Radiative feedback is dominated by ionizing photons and their capacity to reionize the Universe, keeping it ionized until the present epoch. This ionization photoheats the IGM and suppresses a crucial source of cooling: atomic line cooling. Radiative feedback acts to suppress the formation of low-mass galaxies (e.g. \citealt{r86,e92,tw96,ns97,ki00,dhrw}, hereafter DHRW) since the gas is unable to cool and collapse into the dark matter halos to form stars.

Mechanical feedback results from a variety of sources, but up to $z \approx 2$ the main source of this feedback is thought to be galactic outflows caused by many coeval Type II supernovae exploding during the starburst phase. Population III stars may also contribute significantly but early result indicate they will be sub-dominant \citep{nop04}. These outflows do not simply deposit energy into the IGM but also deposit metals. These metals increase the gas cooling efficiency and also provide a convenient observational tracer of this feedback. \citet{sb01} find that most mechanical feedback is caused by low-mass galaxies, however, this is without the inclusion of radiative feedback.

In this work we seek to investigate the degree to which radiative feedback can suppress the formation of low-mass galaxies that form in dark matter halos with virial temperatures $\gtrsim 10^4 {\rm K}$ (below which atomic cooling is inefficient). Hence we do not consider the contribution to our outflows of minihalos and stars which form with a ${\rm H_2}$ cooling mechanism. We do not make specific predictions for the degree of suppression and so the extent of metal enrichment, rather we constrain the potential impact of this effect. We use analytic models for the expansion of galactic outflows described in \cite{pmg07} (hereafter PMG07), which self-consistently include the impact of ram-pressure stripping of halos and metal enrichment. We make a simple parametrisation of the degree of photoionization suppression by choosing  a step-like reionization and using a suppression criterion taken from DHRW.

This paper is set out as follows. In \S\ref{criterion} we describe our criterion for photoionization suppression. In \S\ref{model} we summarize the model of galactic outflows used. The results from our simulations are set out in \S\ref{results}. We discuss the implications of our results and conclude in \S\ref{discconc}.

\section{THE PHOTOIONIZATION SUPPRESSION CRITERION}
\label{criterion}

Our criterion for photoionization suppression can be expressed as follows: halos ionized before turnaround, with a mass below a certain threshold, fail to cool a `sufficient' quantity of gas in order to form a galaxy. This mass threshold corresponds to a particular halo circular velocity $\upsilon_{\rm circ}$ and a particular virial temperature $T_{\rm vir}$.  
We specify this quantity of gas in two ways: 1) $\upsilon_{\rm circ}=\upsilon_{1/2}$, where half the halo baryons or more cool, and 2) $\upsilon_{\rm circ}=\upsilon_0$, where galaxy formation is only suppressed in halos where no gas cools.
The former is an acceptable approximation for our Monte Carlo simulations and is used in our fiducial model. The latter is also considered in the interest of placing a lower limit on the suppression.

\begin{figure}
\centering
\includegraphics[scale=0.45,angle=0]{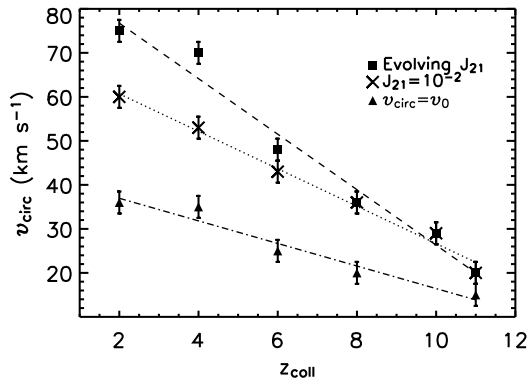}
\caption{The threshold halo circular velocities prevented from forming stars for halo collapse redshift. All points are taken from DHRW. The squares and the crosses show the case where $\upsilon_{\rm circ}=\upsilon_{1/2}$. The squares are for an evolving UV background and the dashed line is a least-squares fit to these points. The crosses show the case were the UV background amplitude has a fixed value of $J_{21}=10^{-2}$ and the dotted line is their fit. The dot-dashed line is the fit to the case where $\upsilon_{\rm circ}=\upsilon_{0}$ and the UV background evolves, given by the triangles.}
\label{vcirc}
\end{figure}

In Figure~\ref{vcirc} we use the results from DHRW's Figure~3 to derive a fit to the evolution of the critical circular velocity for a given collapse redshift, $z_{\rm coll}$. We show their results for three cases: $\upsilon_{1/2}$ for an evolving UV background amplitude ($J=J_{21}\times 10^{-21}$) taken from \citet{ki00}, $\upsilon_{1/2}$ for a fixed amplitude $J_{21}=10^{-2}$, and $\upsilon_0$ for an evolving amplitude. All three have been fitted with the straight lines shown using the least-squares method. Where an evolving UV background is used we find circular velocities,
\begin{equation}
\upsilon_{1/2}=\big[89.4\pm2.4-(6.3\pm0.3)z _{\rm coll}\big] {\rm km\,s^{-1}},
\label{eqv1/2}
\end{equation}
which corresponds to our fiducial model. We also consider modifications to this model using
\begin{equation}
\upsilon_0=\big[42.1\pm2.5-(2.6\pm0.4)z_{\rm coll}\big] {\rm km\,s^{-1}}
\label{eqv0}
\end{equation}
and for a UV background with a fixed amplitude,
\begin{equation}
\upsilon_{1/2}=\big[69.2\pm2.4-(4.3\pm0.3)z_{\rm coll}\big]{\rm km\,s^{-1}}.
\label{eqnoevol}
\end{equation}

These dependences are used in the following work to provide a criterion for photoionization suppression. These circular velocities can be related monotonically in a straightforward manner to a virial temperature, making the criterion simple to incorporate into our cooling time calculation (note that for $T_{\rm vir}=10^4 {\rm K}$, $\upsilon_{\rm circ}\approx 17{\rm km\,s^{-1}}$). We require that the virial temperature is below the value derived for suppression to occur. DHRW argue that the choice of evolving $J_{21}$ constitutes an upper limit on the expected $\upsilon_{\rm circ}$, but it is notable that the effect on mechanical feedback of switching from equation (\ref{eqv1/2}) to equation (\ref{eqnoevol}) are almost indistinguishable in the results that follow.

DHRW perform their simulations using a one-dimensional spherically symmetric hydrodynamics code based on that used by \citet{tw95}. They consider a halo which is crossed by an ionization front during collapse; this occurs at $z=17$ in their simulations. We prefer to choose the crucial reionization redshift, which we call $z_{\rm r}$. We do this in our fiducial model by assuming that, so long as the halo itself is ionized before its turnaround redshift, $z_{\rm turn}$, the precise redshift of reionization is unimportant. This is a reasonable approximation but will underestimate the degree of suppression at high redshift since equations (\ref{eqv1/2})--(\ref{eqnoevol}) are derived from DHRW and for their $z_{\rm coll}=11$, $z_{\rm turn}=18$ and at these redshifts $z_{\rm r}<z_{\rm turn}$. As result we also show the other extreme case, allowing suppression when the $z_{\rm r}>z_{\rm coll}$.

It should be noted that our choice of $z_{\rm r}$ is not intended to be characteristic of reionization as a whole, but rather a good representation of when low-mass halos are ionized. We do not select one value for $z_{\rm r}$ but rather use upper and lower values of $z_{\rm r}=11$ and $z_{\rm r}=6$. The higher redshift value is chosen to correspond to the redshift of reionization derived from the 3-yr WMAP data release in a step-like reionization scenario \citep{page06}. $z_{\rm r}=6$ constitutes a conservative lower limit on the significance of photoionization suppression based on the end of the reionization epoch which may have been detected in quasar absorption spectra (e.g. \citealt{f06,brs06}).

\section{THE MODEL}
\label{model}

For a complete description of the model used see PMG07. We will describe some of the main features of the code and those particularly significant for this work. The approach is a Monte Carlo simulation of a $(12h^{-1}{\rm Mpc})^3$ box, which constitutes a cosmological volume down to the end point of our simulations: $z=2$. A $\Lambda$CDM cosmology consistent with \citet{s06} is used and a Gaussian random realization of structure is formed on a $512^3$ grid. The dark matter halos, their mass, $M$,  and their collapse redshifts, $z_{\rm coll}$, are found. Mergers are included although only in a non-dynamical manner. Upon collapse the gas inside these halos virializes and begins to cool. We use MAPPINGS III\footnote{http://www.mso.anu.edu.au/$\sim$ralph/map.html} \citep{sd93} to calculate the cooling time.

When cooling is complete galaxies are formed with a star formation efficiency $f_*$. Type II supernovae then inject energy into the IGM, preferentially taking a path of least resistance out of large-scale structures with an opening angle, $\alpha$. In our fiducial model we use isotropic outflows but also consider an anisotropic case. The expansion of these outflows out of the gravitational potential well of the host halos is calculated using the thin shell approximation. We assume that the IGM is ionized throughout the expansion of the outflows. This is an acceptable approximation since the epoch of outflow expansion is later than most (if not all) of reionization and so (pre-turnaround) halo suppression.

\section{RESULTS}
\label{results}

\begin{figure*}
\centering
\includegraphics[scale=0.69,angle=90]{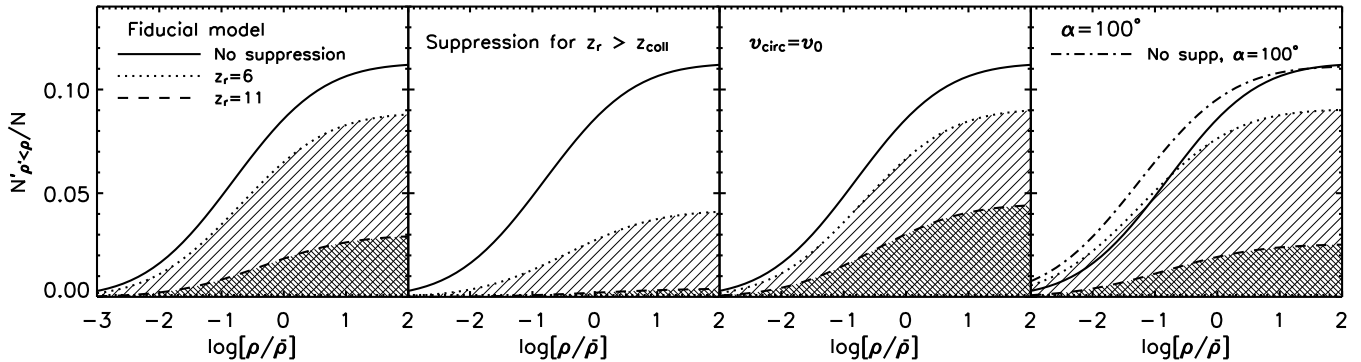}
\caption{The volume filling factor of enriched  regions with a gas overdensity $<\rho/{\bar \rho}$. All panels show the statistic when no photoionization suppression is used in our model with isotropic galactic outflows ({\it solid line}). The {\it first} panel shows the fiducial model of photoionization suppression described in the text for $z_{\rm r}=11$ ({\it dashed line}) and $z_{\rm r}=6$ ({\it dotted line})). The {\it second} and {\it third} panels show the cases where the fiducial model is modified to suppression when $z_{\rm r} > z_{\rm coll}$ and $\upsilon_{\rm circ}=\upsilon_0$ respectively. The {\it fourth} panel shows the case where our fiducial model is modified to include anisotropic outflows with $\alpha=100^\circ$. Here the {\it dot-dashed} line shows the statistic with anisotropic outflows and no suppression.}
\label{enrichstats}
\end{figure*}

We now have the modifications required to perform our simulations of mechanical feedback with a simple description of photoionization suppression of low-mass galaxy formation. We run our simulations to their end point of $z=2$ and compare the regions which have experienced galactic outflows with the underlying gas density field.

As  in PMG07 we describe this IGM density field, $\rho/{\bar \rho}$, by producing a realization of gas structure using Jeans filtering. We map the resultant gas density field to a lognormal probability distribution function to reproduce a degree of non-linear behavior \citep{bd97}. Every point in the $N=512^3$ grid is compared with the outflow locations to determine if it is within the volume of an outflow, the gas density of that point is also noted. In Figure~\ref{enrichstats} we see the results of this statistic. Each panel shows the cumulative statistic ($N^\prime_{\rho^\prime < \rho}/N$) of the number of grid points enriched at a given density $N^\prime_\rho$ , which gives the volume filling factor of enriched systems below this density.  This is shown for the case were no suppression of gas infall is included and the cases where it begins at $z_{\rm r}=11$ and $6$. The second and third panels show hard upper and lower limits (respectively) on our suppression criterion and its impact.

The first panel shows the fiducial case with a suppression criterion using equation~(\ref{eqv1/2}), $z_{\rm turn}<z_{\rm r}$ and isotropic galactic outflows. This shows that photoionization feedback has a pronounced effect at all densities and may result in a $\sim 75\%$ reduction in the volume of the Universe affected by galactic outflows. Even in the conservative case of $z_{\rm r}=6$, a $\sim 25\%$ smaller volume is enriched. This result uses the condition that a galaxy can only be suppressed when its host halo has turned around after it has experienced reionization. If we require instead that suppression may occur even after turnaround and up until collapse, we obtain the results shown in the second panel. This certainly results in excessive suppression.

The third panel shows the case where we use equation~(\ref{eqv0}) to provide the circular velocity for which suppression occurs. This is rather conservative as suppression only occurs where no gas falls into the halo; all other halos are unaffected. The case where $z_{\rm r}=6$ is almost unaffected by this adjustment, while the degree of suppression in the $z_{\rm r}=11$ case is reduced from $\sim 75\%$ to $\sim60\%$.

We show an example of the impact of this suppression for anisotropic outflows in the last panel of Figure~\ref{enrichstats}. This includes a run with an opening angle, $\alpha=100^\circ$ for no suppression, $z_{\rm r}=6$ and $z_{\rm r}=11$. Also shown is the case of isotropic outflows and no suppression for comparison. This illustrates the difference in the nature of the two effects: anisotropic outflows have a strong impact on the degree of enrichment in low density regions (see PMG07) while this suppression affects all densities equally.

\begin{figure*}
\centering
\includegraphics[scale=0.69,angle=90]{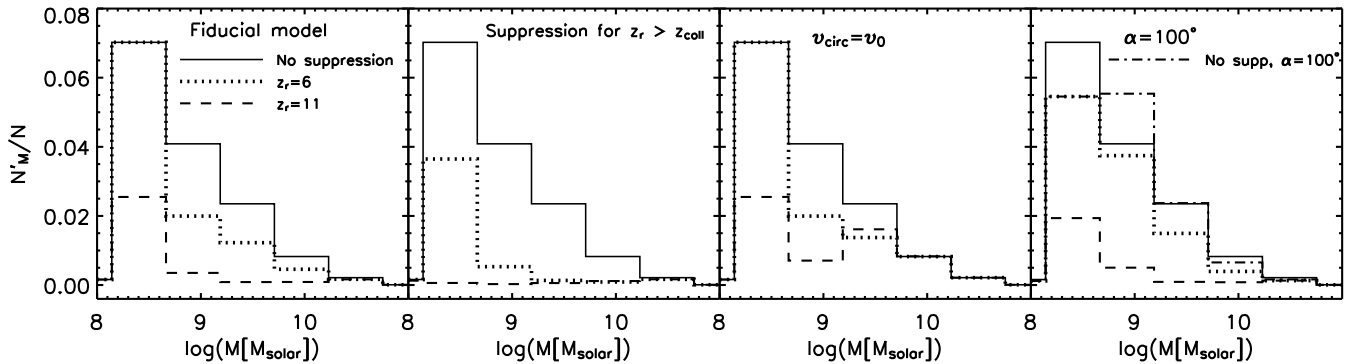}
\caption{The volume filling factor ($N^\prime_M/N$) of regions which have been enriched by galaxies in halos of mass, $M$. The lines correspond to the same models as shown in Figure~\ref{enrichstats}.}
\label{vffmass}
\end{figure*}

Figure~\ref{vffmass} shows the breakdown of volume filling factor enriched by different halo masses for all the models shown in Figure~\ref{enrichstats}. Note that the sum of these histograms is not the total volume filling factor as some regions have been enriched more than once by galaxies of different mass. In all cases the dominant sources of mechanical feedback are galaxies in halos of mass $10^8 {\rm M_\odot} \lesssim M \lesssim 10^9 {\rm M_\odot}$. It is clear that low mass halos dominate the volume filling factor of metals, even when photoionization suppression is included. 

In all but two of the cases considered, the volume enriched peaks around  $10^{8.5} {\rm M_\odot}$ and decreases monotonically with increasing mass. In the first of these two cases, $\upsilon_{\rm circ}=\upsilon_0$ and $z_{\rm r}=11$ are used and the volume enriched rises for halos of mass $\sim10^{9.5} {\rm M_\odot}$. This is mainly because the lower circular velocity threshold allows such halos to escape photoionization suppression. In the second case, anisotropic outflows are used with no suppression resulting in a peak at around $10^{9} {\rm M_\odot}$. This results from a combination of factors related to clustering but mainly: a) a reduction in stripping since outflows take a path of least resistance away from other halos, and b) a lower volume per outflow combined with less overlap in outflows.

It is striking that suppression of gas infall in the $z_{\rm r}=6$ case makes little change to the impact of halos of mass $\lesssim 10^{8.5} {\rm M_\odot}$ in all but the strongest suppression (shown in the second panel). This is because, in a $\Lambda$CDM model, most halos in this mass range turn around early, before $z_{\rm r}$. This also explains why the third panel of Figure~\ref{enrichstats} shows only a small change in the $z_{\rm r}=6$ case compared to the first panel: the significant lowest mass halos are unaffected and so the results are insensitive to the $v_{\rm circ}$ threshold used. The $z_{\rm r}=11$ case shows a significant suppression of galaxies in halos of mass $\lesssim 10^{8.5} {\rm M_\odot}$, however, the impact of suppression on these halos is still small compared to the impact on higher mass halos and they continue to dominate the overall volume filling factor.

\section{DISCUSSION AND CONCLUSIONS}
\label{discconc}

Our analysis of the influence of photoionization suppression of galaxy formation is based on a simple prescription taken from the findings of DHRW. We use an analytic description of galactic outflows in a Monte Carlo cosmological simulation (PMG07). In our fiducial model we have chosen to suppress the formation of galaxies in halos which allow the infall of half their baryons or less. This occurs only when the halo has experienced reionization before its turnaround redshift. These choices, as well as the use of a step-like reionization, are justified simplifications, but in each case we have also bounded the simplification with the lower-limiting case. In all our result we find that this form of radiative feedback significantly reduces the physical extent of enrichment. 

The largest unknown in our prescription for suppression is the reionization redshift. Work must be done to determine what redshift satisfactorily characterizes when halos which dominate the enrichment of the IGM are reionized. This is further complicated interdependence of reionization history and photoionization suppression  (e.g. \citealt{i07}). Also the values taken for $\upsilon_{\rm circ}$ represent upper limits since self-shielding is not included (see DHRW). It may be necessary produce models which describe both radiative transfer and galactic outflows in order to address this issue with any precision. 

We have shown that this suppression has a pronounced impact on the extent of galactic outflows, but one should note that changes in the star formation efficiency result in a similarly pronounced effect \citep{sfm02}. We chose a value of $f_*=0.1$ which is widely accepted (e.g. \citealt{bl00}), but we find that doubling (halving) $ f_*$ leads to $\sim 50\%$ larger (~$\sim 67\%$ smaller) volume filling factor in all our simulations. This change in $f_*$ would of course modify the reionization history (which would counter the impact of changes in $f_*$) but we have not considered this here.

It should be noted that the non-dynamical nature of our models and the way they treat mergers means that we will tend to have more low-mass halos than one would expect. As Figure~\ref{vffmass} shows, one may not simply argue that fewer low-mass halos would result in a weaker suppression signal. The least massive halos which have a significant impact on the IGM by $z=2$ are less affected by this suppression as they are more likely to turn around before reionization. 

Observationally the volume filling factor of the enriched IGM is rather poorly constrained but it appears that the Universe is $\gtrsim5\%$ enriched  \citep{2004MNRAS.347..985P}. If we assume that the star formation efficiency is not significantly higher than our chosen value, our results seem to rule out the strongest photoionization suppression in our models.  Metals have been observed in the IGM at $z\gtrsim6$ \citep{b06} and our results are further indication that metal enrichment and reionization are no-longer two distinct fields but part of a growing field of {\it reionization chemodynamics}.

Using our fiducial model, we find that photoionization suppression may result in a $\sim 75\%$ reduction of the fraction of the Universe enriched at all densities in the IGM. Furthermore, we find that even using a variety of conservative modifications to the suppression, we still find a $\sim 25\%$ reduction in the enrichment of the IGM. We also find that the least massive halos ($\lesssim 10^{8.5} {\rm M_\odot}$) are almost unaffected by late reionization and even with our best reionization redshift we find that these halos are suppressed to a relatively low degree. Despite this suppression, low-mass galaxies continue to dominate the metal enrichment of the Universe by volume.

Further work is required to determine the precise degree of suppression and the largest unknown in our models is the local reionization history experienced by halos. Also a focused study constraining the amount of star formation in halos of various masses which are irradiated at varying stages of collapse would significantly improve our analytic prescription of this effect. Observationally, 30m class telescopes are required to significantly improve constraints on the volume filling factor of metals. In the meantime, more studies must be performed investigating the nature and environment of enriched systems, while extending QSO absorption studies to higher redshift in order to study the evolution of metallicity at the end of the reionization epoch.

\acknowledgments

The authors would like to thank Mark Dijkstra for providing data from DHRW and for his comments on the manuscript. We are also grateful to Ilian Iliev for valuable discussions and Daisuke Kawata for the use of his MAPPINGS III cooling rate table. All calculations were performed at the {\sl Laboratoire
d'astrophysique num\'erique}, Universit\'e Laval.
We thank the Canada Research Chair program and NSERC for support.
%

%

%



\end{document}